\documentstyle[11pt,newpasp,twoside,epsf]{article}
\def\edcomment#1{\iffalse\marginpar{\raggedright\sl#1\/}\else\relax\fi}
\reversemarginpar

\begin{document}
\title{Dark Halo and Disk Galaxy Scaling Laws}
 \author{Julio F. Navarro\altaffilmark{1}}
\affil{Department of Physics and Astronomy, University of Victoria, 3800
Finnerty Road, Victoria, BC, V8P 1A1, Canada}

\altaffiltext{1}{CIAR Scholar, Alfred P.~Sloan Fellow. Email: jfn@uvic.ca}

\begin{abstract}
I highlight recent progress in our understanding of the origin of disk galaxy
scaling laws in a hierarchically clustering universe. Numerical simulations of
galaxy formation in Cold Dark Matter (CDM) dominated universes indicate that the
{\it slope and scatter} of the I-band Tully-Fisher (TF) relation are well
reproduced in this model, although not, as proposed in recent work, because of
the cosmological equivalence between halo mass and circular velocity, but rather
as a result of the dynamical response of the halo to the assembly of the
luminous component of the galaxy. The {\it zero-point} of the TF relation is
determined mainly by the stellar mass-to-light ratio ($\Upsilon_I$) as well as
by the concentration ($c$) of the dark halo. For $c \sim 10$, as is typical of
halos formed in the ``concordance'' $\Lambda$CDM model, we find that this
requires $\Upsilon_I \sim 1.5$, in reasonable agreement with the mass-to-light
ratios expected of stellar populations with colors similar to those of TF
galaxies.  This conclusion supersedes that of Navarro \& Steinmetz (2000a,b),
who claimed the $\Lambda$CDM halos were too concentrated to be consistent with
the observed TF relation. The disagreement can be traced to an incorrect
normalization of the power spectrum used in that work. Our new results show that
simulated disk galaxies in the $\Lambda$CDM scenario are not clearly
inconsistent with the observed I-band Tully-Fisher relation. On the other hand,
their angular momenta is much lower than observed. Accounting simultaneously for
the spin, size and luminosity of disk galaxies remains a challenge for
hierarchical models of galaxy formation.
\end{abstract}

\section{Introduction}

The structural parameters of dark matter halos are tightly related through
simple scaling laws that reflect the cosmological context of their formation.
These regularities are likely the result of the approximately scale-free process
of assembly of collisionless dark matter into collapsed, virialized halos.  One
example is the relation between halo mass and size; a direct result of the
finite age of the universe (see, e.g., Eke, Navarro \& Frenk 1998 and
references within). A second example concerns the angular momentum of dark
halos, which is also linked to mass and size through simple scaling arguments
(Peebles 1969, White 1984, Cole \& Lacey 1996). Finally, similarities are also
apparent in the internal structure of dark halos (Navarro, Frenk \& White 1996,
1997, hereafter NFW).

It has long been thought that the scaling properties of dark halos relate
directly to analogous correlations between structural parameters of disk
galaxies, a question that we have addressed in detail over the past few years
using increasingly sophisticated N-body/gasdynamical simulations (Navarro \&
Steinmetz 1997, Steinmetz \& Navarro 1999, Navarro \& Steinmetz 2000a,b,
hereafter NS00a,b). This work has shown that the velocity scaling of luminosity
and angular momentum in spiral galaxies arise naturally in hierarchical galaxy
formation models such as CDM. Large discrepancies, however, were observed in the
zero-point of both correlations: at fixed rotation speed, simulated disks were
found to be too small and too faint compared with their observational
counterparts.

The failure of simulations to match the angular momentum of disk galaxies was
ascribed to the assembly of the galaxy through a sequence of mergers, where the
bulk of the angular momentum of the gas is transferred to the halo, as first
suggested by Navarro \& Benz (1991). Matching the spin of observed spirals
appears to demand a large injection of energy (presumably from supernovae or
AGNs) that prevents gas at early times from cooling and condensing into
protogalaxies, shifting the bulk of star formation to later times and
alleviating the angular momentum losses associated with major mergers.

The trouble with the zero-point of the Tully-Fisher relation may be traced to
the ``concentration'' of dark halos, which determine the contribution of dark
matter to the circular speed of galaxy disks: the higher the halo concentration
the faster a disk of given mass must rotate to achieve centrifugal
equilibrium. Thus the higher the concentration the lower the stellar
mass-to-light ratio needed for galaxies to remain within the observed
Tully-Fisher relation. As described by NS00, this property can be used to rule
out of the ``standard'' CDM model ($\Omega=1$, $h=0.5$, $\sigma_8=0.6$,
hereafter sCDM): halos formed in this scenario are so concentrated that the
mass-to-light ratio required is unacceptably small.

NS00 also argued that a similar problem afflicts the currently popular
``concordance'' $\Lambda$CDM model ($\Omega_0 \sim0.3$, $\Lambda \sim 0.7$, $h
\sim 0.7$, $\sigma_8 \sim 1$), a result that added to an uncomfortably long list
of concerns regarding the success of CDM on the scale of individual galaxies,
such as the survival of a large number of halos within halos (at odds with the
few satellites observed around the Milky Way; the ``substructure'' problem,
Klypin et al 1999, Moore et al 1999) as well as evidence for constant density
dark matter ``cores'' in some low surface brightness dwarfs (at odds with the
steeply divergent dark matter density profiles expected in CDM universes, see,
e.g., NFW).

Taken together, the evidence appears to warrant a radical revision of one or
more of the premises of the CDM paradigm, and there has been no shortage of
suggestions: self-interacting dark matter (Spergel \& Steinhardt 2000), warm
dark matter (Dalcanton \& Hogan 2000, Bode et al 2000), fluid dark matter
(Peebles 2000), etc., all aim to provide a model that behaves like CDM on large
scales but with reduced substructure and concentration on the scale of
individual galactic halos. Although the introduction of these alternative dark
matter models has generated great interest, it is important to note that the
presumed CDM ``failures'' that motivate them are not beyond doubt. For example,
as noted by van den Bosch et al (2000) and van den Bosch \& Swaters (2000), the
evidence for constant density ``cores'' is sometimes weak and, at best, confined
to a handful of galaxies. At the same time, arguments against the presence of a
large number of ``substructure'' halos in the vicinity of the Milky Way (as CDM
predicts) are indirect and so far inconclusive (see, e.g., White 2000).

I revisit below the argument of NS00 against $\Lambda$CDM based on the large
concentration of dark halos formed in this scenario. As it turns out, the power
spectrum of the $\Lambda$CDM simulations reported by NS00 was incorrectly
normalized: the amplitude of mass fluctuations on $8 h^{-1}$ Mpc scales was
effectively $\sigma_8\approx 1.6$ rather than the quoted $1.14$. Our new
simulations, reported fully in Eke, Navarro \& Steinmetz (2000), show that the
concentration of halos formed in the ``concordance'' $\Lambda$CDM model are not
obviously inconsistent with constraints posed by dynamical observations of the
Milky Way and by the zero-point of the I-band Tully-Fisher relation. I begin
this contribution by reviewing briefly the theoretical motivation for halo
scaling laws and their relation to disk galaxies, and then concentrate on our
new results for the Tully-Fisher relation in the $\Lambda$CDM scenario. I am
grateful to my collaborators, Vincent Eke and Matthias Steinmetz, for allowing
me to discuss these results in advance of publication.

\section{Scaling Laws}

\subsection{Mass, Radius, and Circular Velocity}

The ``size'' of dark halos is usually associated with the distance from the
center at which mass shells are infalling for the first time. This ``virial''
radius (a misnomer, since there is really nothing ``virial'' about it) sets a
firm upper limit to the baryonic mass of the galaxy inside each halo: baryons
beyond this radius have not had time yet to accrete onto the central
galaxy. Virial radii, $r_{\Delta}$, are defined, at $z=0$, by the region that
contains a mean inner density contrast (relative to critical), of order $\Delta
\sim 178 \sqrt{\Omega_0}$. In terms of the circular velocity  at
the virial radius, $V_{\Delta}$, halo masses are given by,
\begin{equation}
M_{\Delta}(V_{\Delta})=1.9 \times 10^{12} \left({\Delta \over 200}\right)^{-1/2}
\left({V_{\Delta} \over 200 \, {\rm km \, s}^{-1}}\right)^3 h^{-1} M_{\odot},
\end{equation}
This power-law dependence on velocity is similar to that of the I-band
Tully-Fisher relation of late-type spirals,
\begin{equation}
L_I \approx 2.0 \times 10^{10} \left({V_{\rm rot} \over 200 \, {\rm km \,
s}^{-1}}\right)^3 h^{-2} L_{\odot},
\end{equation}
a coincidence that suggests a direct cosmological origin for this scaling
law. Introducing the parameters $M_{\rm disk}$ and $\Upsilon_I=M_{\rm disk}/L_I$
to represent the mass of the disk and the disk mass-to-light ratio in solar
units, respectively, eqs.~1 and 2 can be combined to yield $M_{\rm disk}$ as a
fraction of the total mass,
\begin{equation}
f_{\rm mdsk}={M_{\rm disk} \over M_{\Delta}}=8.5 \times 10^{-3} \,
h^{-1}\left({\Delta \over 200}\right)^{1/2} \Upsilon_I \left({V_{\rm rot}\over
V_{\Delta}}\right)^3,
\end{equation}
or, in terms of the total baryonic mass within $r_{\Delta}$ (assuming
$\Omega_b=0.0125 \, h^{-2}$),
\begin{equation}
 f_{\rm bdsk}={M_{\rm disk}\over (\Omega_b/\Omega_0) M_{\Delta}} \approx 0.85 \,
\Omega_0 \, h \, \Upsilon_I \, \left({\Delta \over 200}\right)^{1/2}
\left({V_{\rm rot} \over V_{\Delta}}\right)^3.
\end{equation}
Therefore, the slope and zero-point of the Tully-Fisher relation imply, for a
given cosmogony, a delicate balance between $M_{\rm disk}$, $\Upsilon_I$, and
the ratio $V_{\rm rot}/V_{\Delta}$.

The simplest way to satisfy eqs.~3 and 4 is that argued by Mo, Mao \& White
(1998), who suggest approximately constant values of all these parameters for
all galaxies: $f_{\rm mdsk}\sim 5 \times 10^{-2}$, $\Upsilon
\sim 1.7 h$, and $V_{\rm rot} \sim 1.5 V_{\Delta}$. Although
plausible, this assumption is at odds with numerical experiments, which show
that these parameters vary widely from halo to halo (NS00). This suggests
another possibility: that $f_{\rm mdsk}$, $\Upsilon_I$ and $V_{\rm
rot}/V_{\Delta}$ are not constant from halo to halo but correlated in the manner
prescribed by equation 3. Such correlation may actually arise as a result of the
dynamical response of the dark halo to the assembly of the galaxy.  This is
illustrated in Figure 1a, where we show, for two choices of $\Upsilon_I$, the
relation between $V_{\rm rot}/V_{\Delta}$ and $f_{\rm mdsk}$ computed under the
assumption that the structure of the halo can be approximated by an NFW profile
and that it responds ``adiabatically'' to the assembly of the disk (Mo et al
1998).
\begin{figure}
\plottwo{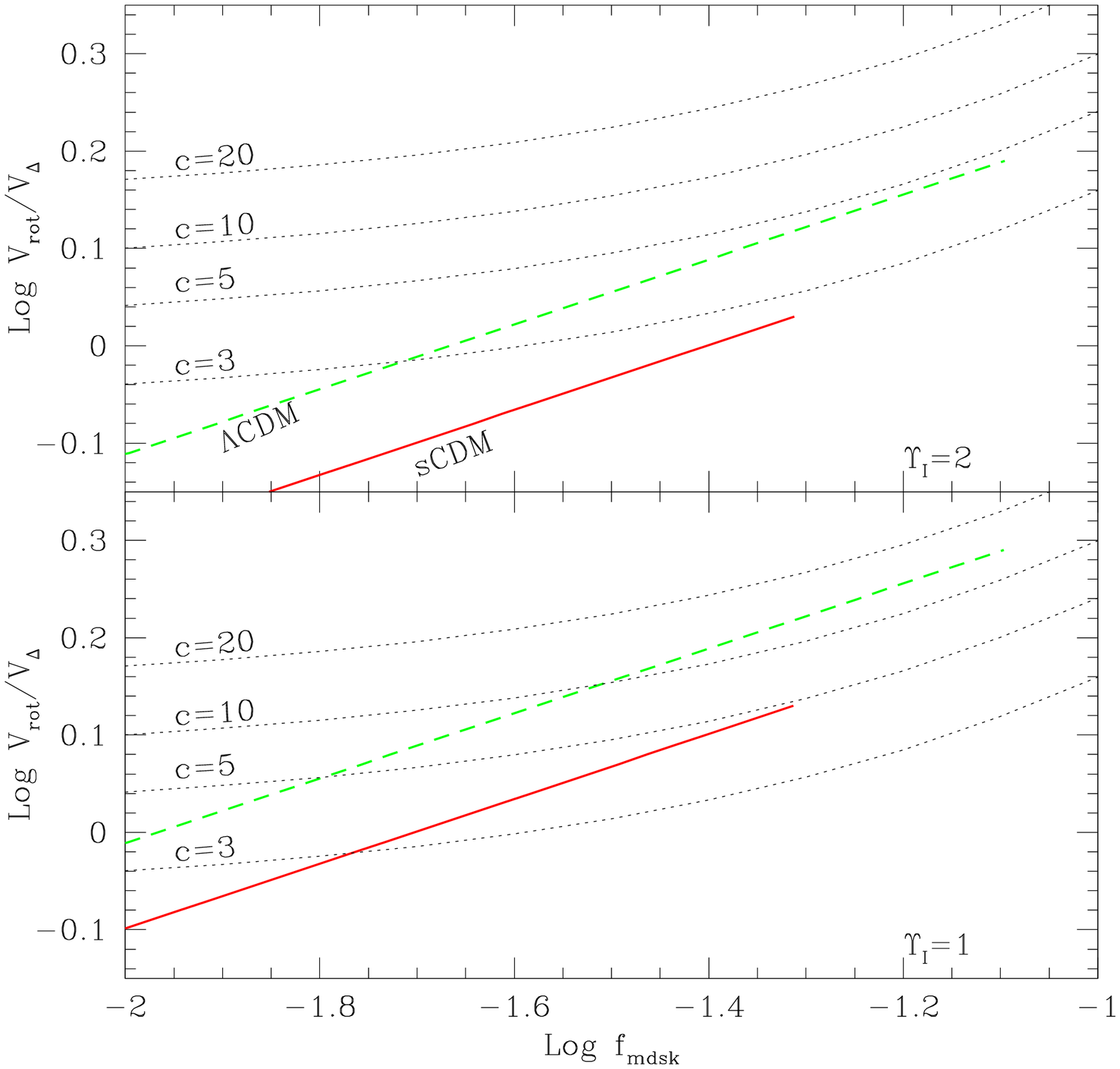}{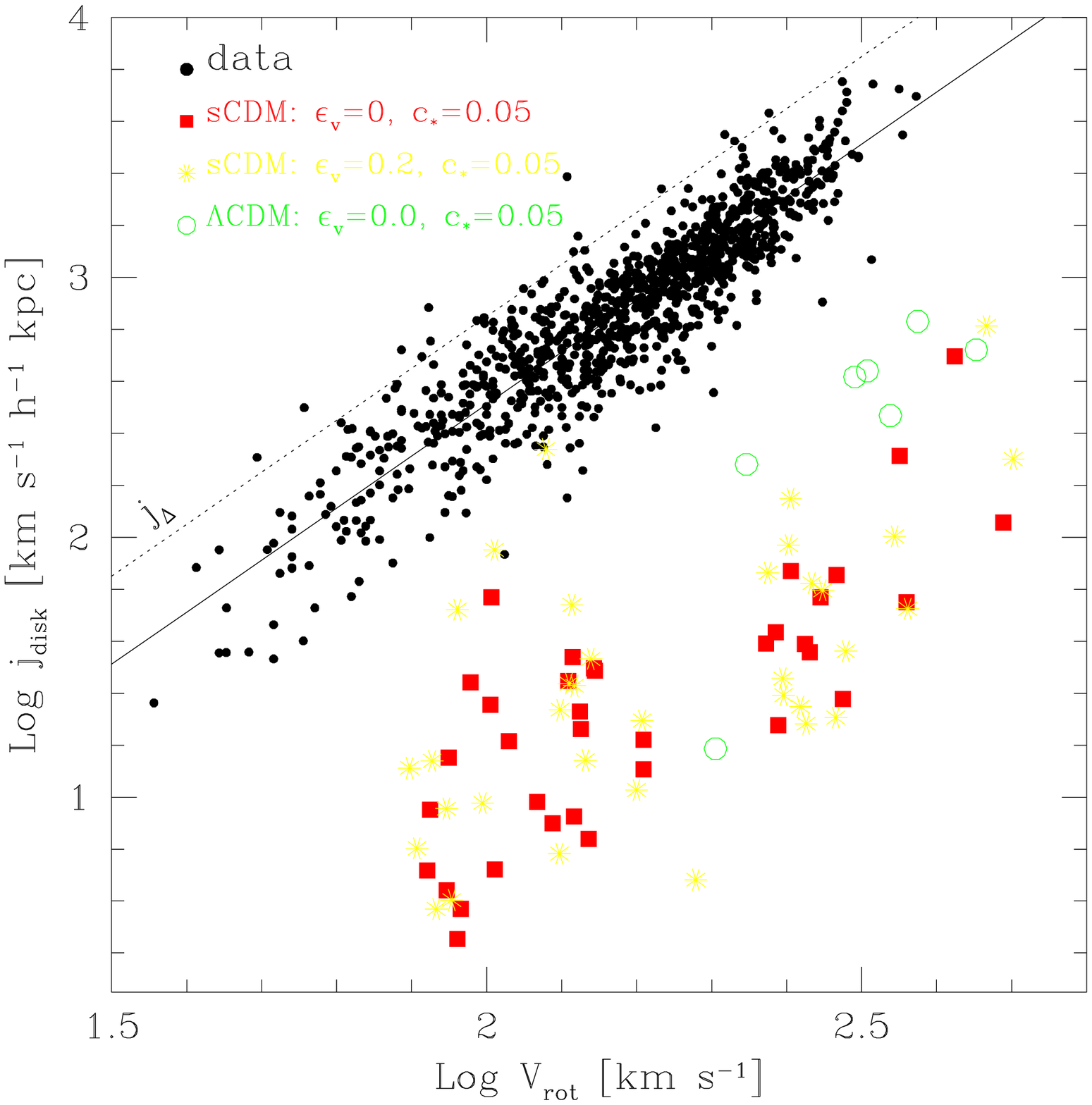}
\caption{{\bf (a) Left panel:}
The disk mass fraction versus the ratio between disk rotation speed and halo
circular velocity.  The thick dashed and solid lines correspond to the {\sl
constraint} imposed on these two quantities by the Tully-Fisher relation (eq.~3)
in the $\Lambda$CDM and sCDM scenarios, respectively.  Dotted lines correspond
to the relation expected for galaxies assembled in NFW halos of constant
``concentration'' parameter, as labeled.  Constant disk mass-to-light ratios and
$\Delta=200$ are assumed throughout; $\Upsilon_I=2$ in the upper panel and
$\Upsilon_I=1$ in the lower one, respectively.  {\bf (b) Right panel:} Specific
angular momentum as a function of circular velocity. Dots are data on TF
galaxies compiled from the literature. Symbols correspond to sCDM and
$\Lambda$CDM models. Two models are shown for sCDM, corresponding to different
choices of the feedback parameters $\epsilon_v$ and $c_*$ (see NS00 for
references and details).}
\end{figure}
The thick solid and dashed lines correspond to the constraint enunciated in
eq.~3 for two different cosmological models; sCDM and $\Lambda$CDM. The
rightmost point in each of the thick lines corresponds to the maximum disk mass
fraction allowed by the baryonic content of the halo. Dotted lines show the
results of applying the adiabatic contraction approximation to the halo for
different values of the NFW ``concentration'' parameter, $c$.
\footnote{$c=r_{\Delta}/r_s$, where $r_s$ is the scale radius of the NFW density
profile, $\rho(r) \propto (r/r_s)^{-1} (1+r/r_s)^{-2}$}
Figure 1a shows that $\Upsilon_I$ and the cosmological parameters determine in
practice the range of halo concentrations consistent with the zero-point of the
TF relation. For example, sCDM halos must have $c \la 5$ if $\Upsilon_I\approx
1$. This effectively rules out the sCDM scenario, since N-body simulations show
that sCDM halos have typically {\it much} higher concentrations, $c\sim 15$-$20$
(NFW). Higher concentrations are acceptable for $\Lambda$CDM, mainly as a result
of the different value of the Hubble constant assumed in that model, which makes
all galaxies dimmer at a given rotation speed. Concentrations as high as $c\sim
10$-$12$ are acceptable if $\Upsilon_I \sim 1$.
%We shall see below that this is in
%agreement with the concentrations obtained in numerical simulations of
%$\Lambda$CDM halo formation.

Another important point illustrated in Figure 1a is that the structure and
dynamical response of the halo to the assembly of the disk may be responsible
for the small scatter in the Tully-Fisher relation. For illustration, consider
two halos of the same mass, and therefore approximately similar concentration,
where the fraction of baryons collected into the central galaxy, $f_{\rm mdsk}$,
differs substantially. Provided that $f_{\rm mdsk} > 0.02$, where the
``adiabatic contraction'' dotted curves are approximately parallel to the
observational constraint delineated by the thick lines, these two galaxies will
lie approximately along the {\it same} Tully-Fisher relation: galaxies scatter
{\it along} the Tully-Fisher relation due to the halo response. Even if the
concentration of the two halos were to differ greatly, its effect on the scatter
of the Tully-Fisher relation would be relatively minor: at fixed $f_{\rm mdsk}$,
$V_{\rm rot}/V_{\Delta}$ changes by only about $20\%$ when $c$ changes by a
factor of two.

\subsection{Circular Velocity and Angular Momentum}

Another similarity between the properties of dark halos and galaxy disks
concerns their angular momentum. N-body simulations show that, in terms of the
dimensionless rotation parameter, $\lambda=J|E|^{1/2}/GM_{\Delta}^{5/2}$ ($J$
and $E$ are the total angular momentum and binding energy of the halo,
respectively), the distribution of halo angular momenta is approximately
independent of mass, redshift, and cosmological parameters, and peaks at around
$\lambda\sim 0.05$ (Cole \& Lacey 1996 and references therein).  The binding
energy depends on the internal structure of the halos but the structural
similarity between dark halos established by NFW implies that $E$ is to good
approximation roughly proportional to $M_{\Delta} V_{\Delta}^2$, with a very
weak dependence on concentration (see Mo et al 1998 for further details). The
specific angular momentum of the halo then may be written as,
\begin{equation}
j_{\Delta}\approx 2 {\lambda \over \Delta^{1/2}} {V_{\Delta}^2\over H_0}= 2.8
\times 10^3 \left({\Delta \over 200}\right)^{-1/2} \left({V_{\Delta} \over 200
\, {\rm km \ s}^{-1}}\right)^2 {\rm km \ s}^{-1} h^{-1} {\rm kpc},
\end{equation}
where we have used the most probable value of $\lambda=0.05$ in the second
equality. The simple velocity-squared scaling of this relation is identical to
that illustrated in Figure 1b between the specific angular momentum of disks and
their rotation speed (solid line in Figure 1b),
\begin{equation}
j_{\rm disk}\approx 1.3 \times 10^3 \left({V_{\rm rot} \over 200\, {\rm km \
s}^{-1}}\right)^2 {\rm km \ s}^{-1} \, h^{-1} {\rm kpc}
\end{equation}
suggestive, as in the case of the Tully-Fisher relation, of a cosmological
origin for this scaling law.

Combining eqs.~5 and 6, we can express the ratio between disk and halo specific
angular momenta as,
\begin{equation}
f_j={j_{\rm disk} \over j_{\Delta}} \approx 0.45 \left({\Delta \over
200}\right)^{1/2} \left({V_{\rm rot}\over V_{\Delta}}\right)^2.
\end{equation}
If the rotation speeds of galaxy disks are approximately the same as
the circular velocity of their surrounding halos, then disks must have
retained about one-half of the available angular momentum during their
assembly. 

The velocity ratio may be eliminated using eq.~4 to obtain a relation between
the fraction of baryons assembled into the disk and the angular momentum ratio,
\begin{equation}
f_j\approx 0.5 \left({\Delta \over 200}\right)^{1/6} 
\left({f_{\rm bdsk}\over\Omega_0 \, h \, \Upsilon_I}\right)^{2/3}. 
\end{equation}
This combined constraint posed by the Tully-Fisher and the angular
momentum-velocity relation is shown in Figure 2a for two different cosmological
models. As in Figure 1a, thick solid lines correspond to the ``standard'' cold
dark matter model, sCDM, and thick dashed lines to the $\Lambda$CDM model. Each
curve is labeled by the value adopted for the disk mass-to-light ratio,
$\Upsilon_I$. The precise values of $f_{\rm bdsk}$ and $f_j$ along each curve
are determined by the ratio $V_{\rm rot}/V_{\Delta}$, and are shown by starred
symbols for the case $V_{\rm rot}=V_{\Delta}$ and $\Upsilon_I=1$.

One important point illustrated by Figure 2a is that disk galaxies formed in a
low-density universe, such as $\Lambda$CDM, need only accrete a small fraction
of the total baryonic mass to match the zero-point of the Tully-Fisher relation,
but must draw a comparably much larger fraction of the available angular
momentum to be consistent with the spins of spiral galaxies. For example, if
$V_{\rm rot}=V_{\Delta}$ and $\Upsilon_I=1$, disk masses amount to only about
$30\%$ of the total baryonic mass of a $\Lambda$CDM halo but contain about
$60\%$ of the available angular momentum. This is intriguing and, at face value,
counterintuitive. Angular momentum is typically concentrated in the outer
regions of the system (see, e.g., Figure 9 in Navarro \& Steinmetz 1997, NS97),
presumably the ones least likely to cool and be accreted into the disk, so it is
puzzling that galaxies manage to tap a large fraction of the available angular
momentum whilst collecting a small fraction of the total mass.  The simulations
in NS97, which include the presence of a strong photo-ionizing UV background,
illustrate exactly this dilemma; the UV background suppresses the cooling of
late-infalling, low-density, high-angular momentum gas and reduces the angular
momentum of cold gaseous disks assembled at the center of dark matter halos.

The situation is less severe in high-density universes such as sCDM; we see from
Figure 2a that disks are required to collect similar fractions of mass and of
angular momentum in order to match simultaneously the Tully-Fisher and the
spin-velocity relations. As a result, any difficulty matching the angular
momentum of disk galaxies in sCDM will become only worse in a low-density
$\Lambda$CDM universe.

Note as well that the problem becomes more severe the lower the mass-to-light
ratio of the disk. Indeed, from the point of view of this constraint, it would
be desirable for disks formed in the $\Lambda$CDM scenario to have $\Upsilon_I >
2$; in this case $f_j \sim f_{\rm bdsk}$ would be consistent with the constraint
posed by observed scaling laws. However, this is the opposite of what is
required to reconcile highly-concentrated halos with the zero-point of the
Tully-Fisher relation. This conundrum illustrates the fact that accounting
simultaneously for the mass and angular momentum of disk galaxies represents a
serious challenge to hierarchical models of galaxy formation.

\section{Numerical Scaling Laws}

\subsection{The Tully-Fisher relation}

The symbols with horizontal error bars in Figure 2b show the numerical
Tully-Fisher relation obtained in our N-body/gasdynamical simulations. These
simulations include the main physical ingredients considered relevant to the
formation of galaxies; self-gravity, hydrodynamical shocks, radiative cooling,
photo-heating, star formation and feedback. Solid (open) circles denote the
luminosities and rotation speeds of galaxy models formed in the $\Lambda$CDM
(sCDM) scenario. Error bars span the range in luminosities corresponding to
assuming either a Salpeter or a Scalo stellar initial mass function. As is clear
from Figure 2b, the slope of the numerical TF relation in both cosmologies is
consistent with the observed values, and the scatter is much smaller than
observed (only $0.12$ mag in the case of $\Lambda$CDM).

A detailed analysis confirms that this is because of the response of the halo to
the assembly of the disk, as discussed in \S2 and in NS00. The main difference
with the latter work concerns the $\Lambda$CDM results: the zero-point of this
relation is offset only by $0.5$ mag, compared with the $1.5$ mag reported by
NS00. The reason for the discrepancy can be traced to the fact that NS00 used an
outdated transfer function for the $\Lambda$CDM transfer function, which led to
a significantly higher effective normalization than intended; instead of
$\sigma_8=1.14$, NS00's simulations had effectively $\sigma_8=1.6$. Halos in a
$\sigma_8=1.6$ universe collapse much earlier and have concentrations about
twice as high as in the ``concordance'' $\sigma_8\approx 1$ $\Lambda$CDM model
(for details see Eke, Navarro \& Steinmetz 2000), leading to a much larger
zero-point offset than seen in Figure 2b.

The remaining $0.5$ mag difference between simulation and observation is perhaps
not too worrying, given that the simulated galaxies have colors that are
slightly too red compared with their TF counterparts. Indeed, the average
$B$-$R$ color in the simulations is $\sim 1.0$, compared with the $\sim 0.8$
average in the sample of Courteau (1997). This suggests that star formation in
the simulations proceeds too quickly and too early; any modification to the
feedback algorithm that remedies this will also tend to make the stellar
population mix in the simulated galaxies brighter. If this correction can bring
$\Upsilon_I$ down by $\sim 50 \%$ then the remaining $0.5$ mag offset between
simulations and observations should be possible to bridge. To summarize, it
appears that, if $\Upsilon \approx 1.5$, then galaxies formed in $\Lambda$CDM
halos have properties that are consistent with the slope, scatter, and
zero-point of the I-band Tully-Fisher relation.

\subsection{The angular momentum problem}

Despite its success at accounting for the TF relation, simulated galaxies fail
to match the observed spin of spiral galaxies in both sCDM and $\Lambda$CDM
scenarios. As in our previous work, we associate these problems with the
assembly of galaxies through merging. The magnitude of the problem is shown in
Figure 1b, where it is clear that the angular momentum of simulated disks is
well below observed values, as a result of the loss of the bulk of their angular
momentum (see symbols in Figure 2a). This serious difficulty may reflect
limitations in our implementation of feedback processes or may indicate a
fundamental flaw in our current hierarchical picture of galaxy
assembly. Nevertheless, it makes clear that accounting simultaneously for the
luminosity, velocity, and angular momentum of spiral galaxies remains a
challenging problem for the popular $\Lambda$CDM cosmogony.

\begin{figure}
\plottwo{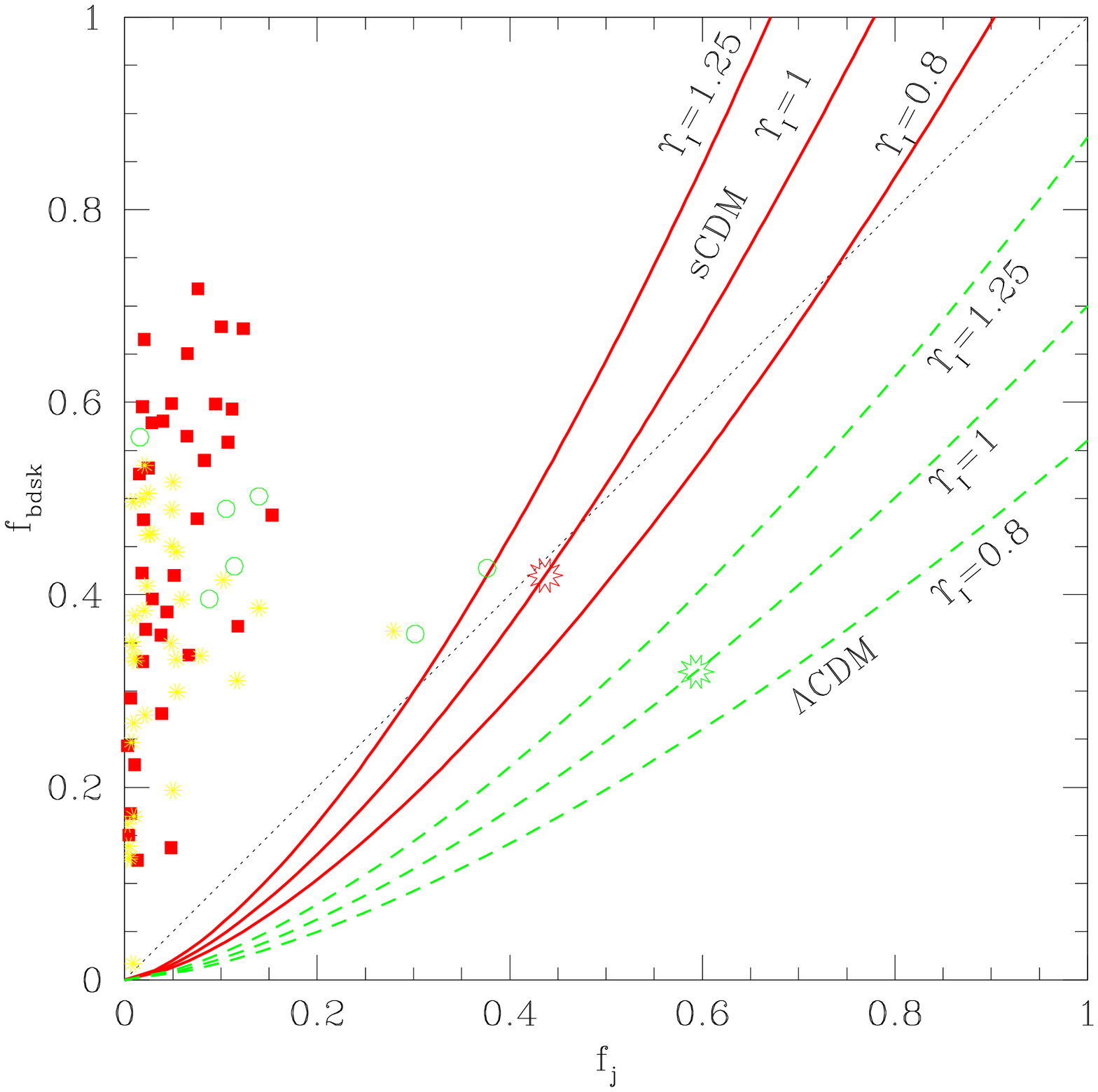}{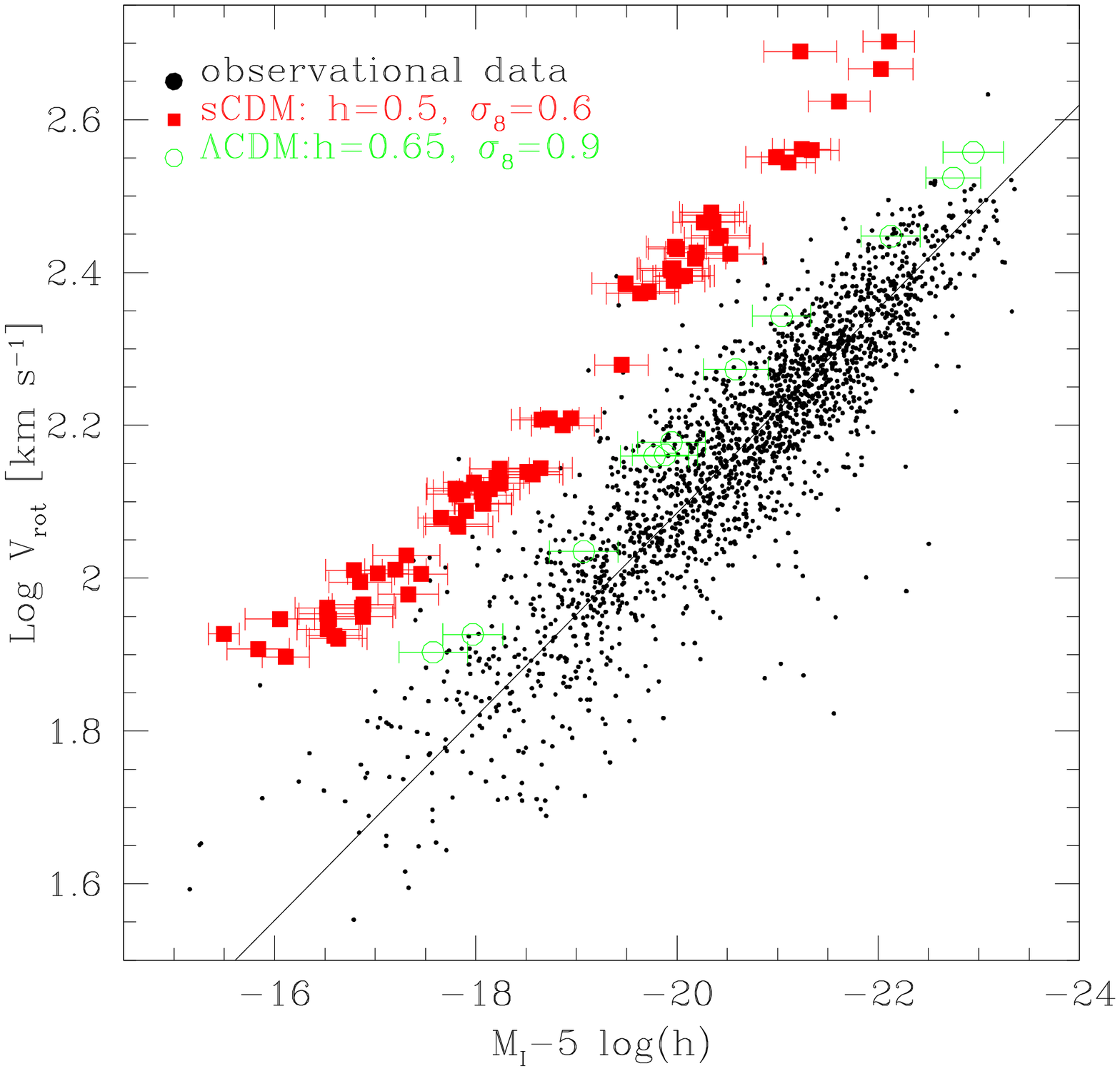}
\caption{{\bf (a) Left panel:}
The fraction of the baryons assembled into a disk galaxy ($f_{\rm bdsk}$) versus
the ratio between the specific angular momenta of the disk and its surrounding
halo ($f_j$). Thick solid and dashed lines correspond to the constraints imposed
by the Tully-Fisher relation and by the relation between rotation speed and
angular momentum (eq.~8). The solid (dashed) thick line corresponds to the sCDM
($\Lambda$CDM) scenario, shown for different values of $\Upsilon_I$, as
labeled. Symbols correspond to simulated galaxy models as per the labels in
Figure 1b. {\bf (b) Right panel:} The I-band Tully-Fisher relation compared with
the results of the numerical simulations. Dots correspond to the observational
samples of Mathewson, Ford \& Buchhorn, (1992), Giovanelli et al (1997), and Han
\& Mould (1992). Error bars in the simulated magnitudes correspond to adopting a
Salpeter or a Scalo IMF. }
\end{figure}
%

%\end{references}
%

\begin{thebibliography}{}

\bibitem[]{} Bode, P., Ostriker, J.P., Turok, N., 2000, (astro-ph/0010389).

\bibitem[]{} Cole S., \& Lacey C.\ 1996, MNRAS, 281, 716

\bibitem[]{Cou97} Courteau, S.\ 1997, \aj, 114, 2402

\bibitem[]{} Dalcanton, J.J., Hogan, C. \ 2000, \apj, submitted (astro-ph/0004381).

\bibitem[]{} Eke, V., Navarro, J.F., Frenk, C.S.\ 1998, \apj, 503, 569

\bibitem[]{} Eke, V., Navarro, J.F., Steinmetz, M. \ 2000, \apj, submitted.

\bibitem[]{gio97} Giovanelli, R., Haynes, M.P., Herter, T., \& Vogt, N.P.\ 1997,
\aj, 113, 22

\bibitem[]{} Han, M., \& Mould, J.R.\ 1992, \apj, 396, 453

\bibitem[]{} Klypin, A., Kravtsov, A., Valenzuela, O., \& Prada, F. \ 1999, \apj, 522,
82.

\bibitem[]{} Mathewson, D.S., Ford, V.L., \& Buchhorn, M.\ 1992, \apjs, 81, 413

\bibitem[]{} Mo, H.J., Mao, S.,  White, S.D.M.\ 1998, MNRAS, 295, 319

\bibitem[]{} Moore, B., et al. \ 1999, \apj, 524, L19.

\bibitem[]{} Navarro, J.F., Benz, W., 1991, \apj, 380, 320

\bibitem[]{nfw96} Navarro, J.F., Frenk, C.S., White, S.D.M.\ 1996, \apj, 462,
563 (NFW)

\bibitem[]{nfw97} Navarro, J.F., Frenk, C.S., White, S.D.M.\ 1997, \apj, 490,
493 (NFW)

\bibitem[]{nav97} Navarro, J.F., Steinmetz, M.\ 1997, \apj, 478, 13 (NS97)

\bibitem[]{nav00} Navarro, J.F., Steinmetz, M.\ 2000a, \apj, 528, 607 (NS00a)

\bibitem[]{nav00} Navarro, J.F., Steinmetz, M.\ 2000b, \apj, 538, 477 (NS00b).

\bibitem[]{} Peebles, P.J.E., 1969, \apj, 155, 393 

\bibitem[]{} Peebles, P.J.E., 2000, \apj, 534, L127.

\bibitem[]{} Spergel, D., Steinhardt, P. \ 2000, \prl, 84, 3760.

\bibitem[]{ste99} Steinmetz, M., Navarro, J.F.\ 1999, \apj,  513, 555


\bibitem[]{} van den Bosch, Swaters, R. \ 2000, \aj, submitted (astro-ph/0006048)

\bibitem[]{} van den Bosch, F.C., Robertson, B.E., Dalcanton, J.J., de Blok,
W.J.G. \ 2000, \aj, 119, 1579.

\bibitem[]{} White, S.D.M. 1984, \apj, 286, 38

\bibitem[]{} White, S.D.M. 2000, at the ITP Conference on Galaxy Formation (\verb"http://online.itp.ucsb.edu/online/galaxy_c00/white/")

\end{thebibliography}
\end{document}